\documentclass[aps,prd,superscriptaddress,notitlepage,nofootinbib,twocolumn]{revtex4-2}
\usepackage[T1]{fontenc}
\usepackage[utf8]{inputenc}
\usepackage{amsmath,amssymb,mathtools,microtype,cancel}
\usepackage[hidelinks]{hyperref}
\DeclareMathOperator{\trace}{tr}
\DeclareMathOperator{\sign}{sgn}

\begin{document}
\title{Hamiltonian Analysis of Pre-geometric Gravity}

\author{Andrea Addazi}
    \email{addazi@scu.edu.cn}
    \affiliation{Center for Theoretical Physics, College of Physics Science and Technology, Sichuan University, 610065 Chengdu, China}
    \affiliation{Laboratori Nazionali di Frascati INFN, Frascati (Rome), Italy, EU}
\author{Salvatore Capozziello}
    \email{capozziello@na.infn.it}
    \affiliation{Dipartimento di Fisica ``E.\ Pancini'', Università di Napoli ``Federico II'', Via Cinthia 9, 80126 Napoli, Italy}
    \affiliation{Scuola Superiore Meridionale, Largo San Marcellino, 10, 80138, Napoli, Italy}
    \affiliation{INFN Sezione di Napoli, Complesso Universitario di Monte Sant'Angelo, Edificio 6, Via Cintia, 80126, Napoli, Italy}
    \affiliation{\mbox{Research Center of Astrophysics and Cosmology, Khazar University, Baku, AZ1096, 41 Mehseti Street, Azerbaijan}}
\author{Antonino Marcianò}
    \email{marciano@fudan.edu.cn}
    \affiliation{Center for Astronomy and Astrophysics, Center for Field Theory and Particle Physics, and Department of Physics, Fudan University, Shanghai 200438, China}
    \affiliation{Laboratori Nazionali di Frascati INFN, Frascati (Rome), Italy, EU}
    \affiliation{INFN sezione di Roma ``Tor Vergata'', 00133 Rome, Italy, EU}
\author{Giuseppe Meluccio}
    \email{giuseppe.meluccio-ssm@unina.it}
    \affiliation{Scuola Superiore Meridionale, Largo San Marcellino, 10, 80138, Napoli, Italy}
    \affiliation{INFN Sezione di Napoli, Complesso Universitario di Monte Sant'Angelo, Edificio 6, Via Cintia, 80126, Napoli, Italy}
\date{\today}

\begin{abstract}
    The Einstein--Cartan theory of gravity can arise from a mechanism of spontaneous symmetry breaking within the context of pre-geometric gauge theories. In this work, we develop the Hamiltonian analysis of such theories. By making contact with the ADM formalism, we show that all the results of canonical General Relativity are correctly recovered in the IR limit of the spontaneously broken phase. We then apply Dirac's algorithm to study the algebra of constraints and determine the number of degrees of freedom in the UV limit of the unbroken phase. We also discuss possible pathways toward a UV completion of General Relativity, including a pre-geometric generalisation of the Wheeler--DeWitt equation and an extended BF formulation of the pre-geometric theory.
\end{abstract}

\maketitle

\def\sectionautorefname{Appendix}

\section{Introduction}
As a result of the spontaneous symmetry breaking (SSB) of a fundamental gauge symmetry of spacetime, a classical metric structure can emerge out of a pre-geometric four-dimensional spacetime. This can be achieved in a gauge theory {\it à la} Yang--Mills, in which neither the spacetime metric nor the tetrads are assumed to exist, corresponding to an unbroken phase \cite{addazi:pre-geometry}. The pre-geometric field content of this formulation consists of the gauge field $A_\mu^{AB}$ of the group $SO(1,4)$ or $SO(3,2)$ and a Higgs-like scalar field $\phi^A$. Once the latter acquires a nonzero vacuum expectation value (v.e.v.) because of its dynamics, the ensuing SSB reduces the gauge group of spacetime to the Lorentz group $SO(1,3)$, in compliance with the Einstein equivalence principle. This, in turn, allows recovering the Einstein--Cartan theory with a cosmological constant term after the identification of the spin connection $\omega_\mu^{ab}$ and the tetrads $e_\mu^a$ of the gravitational field \cite{addazi:pre-geometry,randono:gauge,westman:gauge,westman:cartan,westman:introduction}. There exist two possible Lagrangian densities which can realise the emergence of Einstein gravity from a pre-geometric framework: the one proposed by {\it Wilczek} \cite{wilczek:gauge} and the one proposed by {\it MacDowell} \& {\it Mansouri} \cite{macdowell:unified}. With a view to explore a novel path towards a theory of Quantum Gravity, in this work we address the Hamiltonian analysis of such proposals; in fact, this is key to studying the quantisation and the UV limit of the gravitational interaction. We also discuss a connection with BF theories, which can provide interesting tools for understanding gravity in the pre-geometric picture.

The paper is organised as follows. In Sec.\ \ref{sec:2} we present the canonical analysis of the Wilczek theory, which in Sec.\ \ref{sec:4} is extended by using Dirac's algorithm for gauge theories. In Sec.\ \ref{sec:3} we make contact with the Hamiltonian analysis of the Einstein--Cartan gravity in the Arnowitt--Deser--Misner (ADM) formalism, showing how it is recovered in the spontaneously broken phase of the Wilczek theory. Sec.\ \ref{sec:5} contains a brief discussion of the generalisation of the Wheeler--DeWitt equation in the pre-geometric framework under consideration. Before drawing our conclusions in Sec.\ \ref{sec:7}, we present, in Sec.\ \ref{sec:6}, a formulation of the pre-geometric theory in terms of an extension of a topological BF theory. In the \autoref{sec:appendix} we then reproduce the canonical analysis for the MacDowell--Mansouri theory. We adopt natural units along the paper.

\section{The canonical analysis}\label{sec:2}
The Lagrangian density that we will focus on in this section is the sum of the Wilczek Lagrangian density \cite{wilczek:gauge} and a potential term for the realisation\footnote{Compared to the paper \cite{addazi:pre-geometry}, the kinetic terms for the pre-geometric fields will be excluded from this analysis for the sake of simplicity. Another reason to neglect these terms is that they are not essential except for the study of the Higgs mechanism associated with the SSB of the fundamental gauge symmetry of spacetime.}  of the SSB \cite{addazi:pre-geometry}, respectively
\begin{subequations}
    \begin{align}
        \mathcal{L}_\textup{W}&=k_\textup{W}\epsilon_{ABCDE}\epsilon^{\mu\nu\rho\sigma}F_{\mu\nu}^{AB}\nabla_\rho\phi^C\nabla_\sigma\phi^D\phi^E,\\
        \mathcal{L}_\textup{SSB}&=-k_\textup{SSB}v^{-4}\lvert J\rvert(\phi^2\mp v^2)^2,\qquad\phi\equiv\sqrt{\eta_{AB}\phi^A\phi^B},\label{eq:potential}
    \end{align}
\end{subequations}
where $F_{\mu\nu}^{AB}$ is the field strength of the connection $A_\mu^{AB}$, $\nabla_\mu$ denotes the gauge covariant derivative and $J$ is a pre-geometric quantity:
\begin{align*}
    F_{\mu\nu}^{AB}&=2\partial_{[\mu}A_{\nu]}^{AB}+2A_{C[\mu}^AA_{\nu]}^{CB},\\
    \nabla_\lambda\phi^A&=\partial_\lambda\phi^A+A_{B\lambda}^A\phi^B,\\
    J&\equiv\epsilon_{ABCDE}\epsilon^{\mu\nu\rho\sigma}\nabla_\mu\phi^A\nabla_\nu\phi^B\nabla_\rho\phi^C\nabla_\sigma\phi^D\phi^E,
\end{align*}
with $A^A_{B\mu}\equiv\eta_{BC}A_\mu^{AC}$, $k_\textup{W}$ and $v$ nonzero constants and $k_\textup{SSB}$ a positive constant. The v.e.v.\ of the field $\phi^A$ is $v\delta^A_5$. Uppercase Latin indices $A,B,C$ etc.\ run from $1$ to $5$, lowercase Latin indices $a,b,c$ etc.\ and Greek indices $\lambda,\mu,\nu$ etc.\ run from $0$ to $3$, Greek indices $\alpha,\beta,\gamma$ etc.\ and lowercase Latin indices $i,j,k$ etc.\ run from $1$ to $3$. The internal space metric for the tangent spaces to the spacetime manifold is a generalised Minkowski metric $\eta$ with signature $(-,+,+,+,+)$ or $(+,+,+,-,-)$ depending on the gauge group under consideration,\footnote{In what follows, whenever a double sign is encountered, the upper one will refer to the case of $SO(1,4)$, while the lower one to $SO(3,2)$.} respectively $SO(1,4)$ or $SO(3,2)$. The Hamiltonian analysis will then be reproduced for the MacDowell--Mansouri Lagrangian density \cite{macdowell:unified} in the \autoref{sec:appendix}, following the same steps as for the case of the Wilczek Lagrangian density.

Observing that
\begin{align}
    \frac{\delta J}{\delta\dot{\phi}^E}&=4\epsilon_{ABCDE}\epsilon^{0ijk}\nabla_i\phi^A\nabla_j\phi^B\nabla_k\phi^C\phi^D,\label{eq:delta-J}\\
    \frac{\delta F_{\mu\nu}^{AB}}{\delta\dot{A}_\lambda^{CD}}&=2\delta_{[\mu}^0\delta_{\nu]}^\lambda\delta_{[C}^A\delta_{D]}^B,
\end{align}
the conjugate momenta to the pre-geometric fields $\phi^A$ and $A_\mu^{AB}$ can be computed respectively as
\begin{align}\label{eq:mom_phi}
    \begin{split}
        \Pi_E&=\frac{\delta(\mathcal{L}_\textup{W}+\mathcal{L}_\textup{SSB})}{\delta\dot{\phi}^E}=2\epsilon_{ABCDE}\epsilon^{0ijk}\nabla_k\phi^C\phi^D[k_\textup{W}F_{ij}^{AB}\\
        &-2\sign(J)k_\textup{SSB}v^{-4}\nabla_i\phi^A\nabla_j\phi^B(\phi^2\mp v^2)^2],
    \end{split}\\
    \Pi^\lambda_{DE}&=\frac{\delta(\mathcal{L}_\textup{W}+\mathcal{L}_\textup{SSB})}{\delta\dot{A}_\lambda^{DE}}=2k_\textup{W}\epsilon_{ABCDE}\epsilon^{0\lambda jk}\nabla_j\phi^A\nabla_k\phi^B\phi^C\nonumber;
\end{align}
in particular,
\begin{subequations}\label{eq:mom_A}
    \begin{align}
        \Pi^i_{DE}&=2k_\textup{W}\epsilon_{ABCDE}\epsilon^{0ijk}\nabla_j\phi^A\nabla_k\phi^B\phi^C,\\
        \Pi^0_{DE}&=0.
    \end{align}
\end{subequations}
In terms of the conjugate momenta, the total Lagrangian density can be recast as
\begin{equation}\label{eq:lagrangian}
    \mathcal{L}_\textup{W}+\mathcal{L}_\textup{SSB}=\Pi^i_{AB}F_{0i}^{AB}+\Pi_A\nabla_0\phi^A.
\end{equation}
The corresponding Hamiltonian density does not depend on the time derivatives of the pre-geometric fields:
\begin{equation}\label{eq:H-density_W}
    \begin{split}
        \mathcal{H}_\textup{W}&=\Pi_{AB}^i\dot{A}_i^{AB}+\Pi_A\dot{\phi}^A-\mathcal{L}_\textup{W}-\mathcal{L}_\textup{SSB}\\
        &=\Pi_{AB}^i(\partial_iA_0^{AB}-2A_{C[0}^AA_{i]}^{CB})-\Pi_AA_{B0}^A\phi^B.
    \end{split}
\end{equation}
Performing an integration by parts and retaining the boundary term, the Hamiltonian is
\begin{equation}\label{eq:H_W(unbroken)}
    \begin{split}
        H_\textup{W}&=\int\mathcal{H}_\textup{W}\,d^3x=\int[-A_0^{AB}\partial_i\Pi^i_{AB}+\partial_i(\Pi^i_{AB}A_0^{AB})]\,d^3x\\
        &+\int(-2\Pi^i_{AB}A_{C[0}^AA_{i]}^{CB}-\Pi_AA_{B0}^A\phi^B)\,d^3x\\
        &=-\int A_0^{AB}(\partial_i\Pi^i_{AB}+2\Pi^i_{BC}A_{Ai}^C+\eta_{BC}\Pi_A\phi^C)\,d^3x\\
        &+\int\partial_i(\Pi^i_{AB}A_0^{AB})\,d^3x,
    \end{split}
\end{equation}
where in the last step we made use of the identity
\begin{equation*}
    \Pi^i_{AB}A_{C[0}^AA_{i]}^{CB}=\Pi^i_{BC}A_0^{AB}A_{Ai}^C.
\end{equation*}

\section{Recovering the Einstein--Cartan gravity in the ADM formalism}\label{sec:3}
The SSB of the vacuum state in the Wilczek theory of emergent gravity yields the following results (see Ref.\ \cite{addazi:pre-geometry} for the details):
\begin{equation}\label{eq:SSB}
    \begin{split}
        \phi^A\xrightarrow{SSB}v\delta_5^A,\qquad A_\mu^{ab}\xrightarrow{SSB}\omega_\mu^{ab}&,\qquad A_\mu^{a5}\xrightarrow{SSB}me_\mu^a,\\
        \mathcal{L}_\textup{W}\xrightarrow{SSB}k_\textup{W}v^3m^2\epsilon_{abcd}\epsilon^{\mu\nu\rho\sigma}e_\mu^a&e_\nu^bR_{\rho\sigma}^{cd}\pm48k_\textup{W}v^3m^4e,\\
        \mathcal{L}_\textup{SSB}\xrightarrow{SSB}0,\qquad J&\xrightarrow{SSB}-24v^5m^4e,\\
        F_{\mu\nu}^{AB}\xrightarrow{SSB}R_{\mu\nu}^{ab}\mp2m^2e^a_{[\mu}e^b_{\nu]},&\quad\nabla_\lambda\phi^A\xrightarrow{SSB}\pm vme^a_\lambda,
    \end{split}
\end{equation}
where $m$ is a nonzero constant, $e$ is the tetrad determinant and the Riemann tensor is defined as
\begin{equation*}
    R_{\mu\nu}^{ab}=2\partial_{[\mu}\omega_{\nu]}^{ab}+2\omega^a_{c[\mu}\omega^{cb}_{\nu]}.
\end{equation*}
For later convenience, we can introduce other two auxiliary quantities \cite{addazi:pre-geometry}, namely 
\begin{equation*}
    \begin{split}
        P_{\mu\nu}&\equiv\eta_{AB}\nabla_\mu\phi^A\nabla_\nu\phi^B,\\
        w^\mu_A&\equiv\pm\epsilon_{ABCDE}\epsilon^{\mu\nu\rho\sigma}\nabla_\nu\phi^B\nabla_\rho\phi^C\nabla_\sigma\phi^D\phi^E,
    \end{split}
\end{equation*}
their SSB being provided respectively by
\begin{equation}\label{eq:SSB_aux}
    \begin{split}
        P_{\mu\nu}&\xrightarrow{SSB}v^2m^2\eta_{ab}e_\mu^ae_\nu^b=v^2m^2g_{\mu\nu},\\
        w^\mu_A&\xrightarrow{SSB}v^4m^3\epsilon_{abcd}\epsilon^{\mu\nu\rho\sigma}e_\nu^be_\rho^ce_\sigma^d=-6v^4m^3ee^\mu_a.
    \end{split}
\end{equation}
By identifying the reduced Planck mass and the cosmological constant respectively as
\begin{subequations}
    \begin{align}
        M_\textup{P}^2&\equiv-8k_\textup{W}v^3m^2,\label{eq:M_P^2}\\
        \Lambda&\equiv\pm6m^2=\mp\frac{3M_\textup{P}^2}{4k_\textup{W}v^3},\label{eq:lambda}
    \end{align}
\end{subequations}
the SSB of the Wilczek Lagrangian density can be seen to reproduce the gravitational Lagrangian density in the Palatini formalism, that is
\begin{equation*}
    \mathcal{L}_\textup{W}\xrightarrow{SSB}\mathcal{L}_\textup{EH}+\mathcal{L}_\Lambda,
\end{equation*}
where the Einstein--Hilbert Lagrangian density and the cosmological constant term are defined respectively as
\begin{equation*}
    \mathcal{L}_\textup{EH}=\frac{M_\textup{P}^2}{2}ee^\mu_ae^\nu_bR^{ab}_{\mu\nu},\qquad\mathcal{L}_\Lambda=-M_\textup{P}^2\Lambda e,
\end{equation*}
with $e\equiv\det(e_\mu^a)$. This result is made possible by exploiting the identity
\begin{equation*}
    \epsilon_{abcd}\epsilon^{\mu\nu\rho\sigma}e_\mu^ae_\nu^bR_{\rho\sigma}^{cd}=-4ee^\mu_ae^\nu_bR^{ab}_{\mu\nu},
\end{equation*}
which holds true only if the tetrads are assumed to be invertible \cite{addazi:pre-geometry}. After the SSB, the conjugate momenta \eqref{eq:mom_phi} and \eqref{eq:mom_A} become respectively
\begin{equation}\label{eq:SSB_phi}
    \Pi_A\xrightarrow{SSB}\Pi_a=\pm2k_\textup{W}v^2m\epsilon_{abcd}\epsilon^{0ijk}e^d_k(R^{bc}_{ij}\mp2m^2e^b_ie^c_j)
\end{equation}
and
\begin{subequations}
    \begin{align}
        \Pi^i_{AB}\xrightarrow{SSB}\Pi^i_{ab}&=2k_\textup{W}v^3m^2\epsilon_{abcd}\epsilon^{0ijk}e^c_je^d_k,\label{eq:SSB_A}\\
        \Pi^0_{AB}\xrightarrow{SSB}\Pi^0_{ab}&=0.
    \end{align}
\end{subequations}
Note that Eqs.\ \eqref{eq:SSB}, \eqref{eq:SSB_phi} and \eqref{eq:SSB_A} imply that
\begin{equation}\label{eq:pqs}
    \begin{split}
        \Pi^i_{AB}\dot{A}_i^{AB}&+\Pi_A\dot{\phi}^A=\Pi^i_{ab}\dot{A}_i^{ab}+2\Pi^i_{a5}\dot{A}_i^{a5}+\Pi_A\dot{\phi}^A\\
        &\xrightarrow{SSB}\Pi^i_{ab}\dot{\omega}_i^{ab}\cancel{+2m\Pi^i_{a5}\dot{e}_i^a}\cancel{+\dot{v}\Pi_A\delta^A_5}.
    \end{split}
\end{equation}
This result shows that the (spatial) spin connection plays a privileged role in the Hamiltonian formulation of the theory, as opposed to the (spatial) tetrads and the Higgs-like scalar field. Finally, the SSB of the Hamiltonian \eqref{eq:H_W(unbroken)} is given by
\begin{equation}\label{eq:H_W(broken)}
    \begin{split}
        H_\textup{W}&\xrightarrow{SSB}-\int(\omega_0^{ab}\partial_i\Pi^i_{ab}+2\Pi^i_{ab}\omega_{c0}^a\omega_i^{cb})\,d^3x\\
        &+\int[\pm2m^2\Pi^i_{ab}e^a_0e^b_i\mp vm\Pi_ae^a_0+\partial_i(\Pi^i_{ab}\omega_0^{ab})]\,d^3x\\
        &=\frac{M_\textup{P}^2}{4}\epsilon_{abcd}\epsilon^{0ijk}\biggl\{\int[\omega_0^{ab}\partial_i(e^c_je^d_k)+2\omega_{e0}^a\omega_i^{eb}e^c_je^d_k]\,d^3x\\
        &+\int[\mp4m^2e^a_0e^b_ie^c_je^d_k+e^a_0e^b_iR^{cd}_{jk}-\partial_i(\omega_0^{ab}e^c_je^d_k)]\,d^3x\biggr\}.
    \end{split}
\end{equation}
Observe that the Hamiltonian $H_\textup{W}$ is polynomial in the fields and their conjugate momenta, as well as their derivatives, in both the unbroken \eqref{eq:H_W(unbroken)} and spontaneously broken \eqref{eq:H_W(broken)} phases -- this is a consequence of the respective Lagrangian densities being polynomial as well.

The canonical Hamiltonian formulation of General Relativity (GR) is obtained by means of the ADM formalism \cite{arnowitt:dynamical}, which takes advantage of a foliation of spacetime into three-dimensional spatial hypersurfaces parametrised by a time coordinate. The resulting spacetime metric and its inverse can be expressed as
\begin{equation*}
    \begin{split}
        g_{\mu\nu}&=\begin{pmatrix}
                        N_iN^i-N^2&N_j\\
                        N_i&h_{ij}
                    \end{pmatrix},\\
        g^{\mu\nu}&=\frac{1}{N^2}\begin{pmatrix}
                                     -1&N^j\\
                                     N^i&N^2h^{ij}-N^iN^j
                                 \end{pmatrix},
    \end{split}
\end{equation*}
where $N$ is the lapse function, $N^i$ are the shift functions, $h_{ij}$ is the induced metric on the spatial hypersurfaces and $h^{ij}$ is its inverse, with $h^{ik}h_{kj}=\delta^i_j$ and $N_i=h_{ij}N^j$. The determinant $g\equiv\det(g_{\mu\nu})$ of the four-dimensional metric and that $h\equiv\det(h_{ij})$ of the three-dimensional metric are related by the formula $\sqrt{-g}=N\sqrt{h}$. The spatial metric in each tangent space to the spatial hypersurfaces is the Euclidean metric $\delta$ with signature $(+,+,+)$. The normal vector to the spatial hypersurfaces has components
\begin{equation*}
    n_\mu=-N\delta_\mu^0,\qquad n^\mu=\frac{1}{N}(1,-N^i),
\end{equation*}
hence it is normalised as $n_\mu n^\mu=-1$. The introduction of the tetrads via the relation $g_{\mu\nu}=e_\mu^ae_\nu^b\eta_{ab}$ allows not only to translate the second-order formalism of GR into the first-order one of the Einstein--Cartan theory, but also to conveniently recast the ADM formulation in terms of local inertial frames. For instance, the unit normal can be recast as $n_a=e^\mu_an_\mu=-Ne^0_a$, from which follows that
\begin{equation}\label{eq:e0a-ea0}
    e^0_a=-\frac{n_a}{N},\qquad e^a_0=Nn^a+N^ie_i^a.
\end{equation}
In addition, the relation between the metric determinants can be restated in terms of the tetrad determinants as $e=N\prescript{(3)}{}{e}$, where
\begin{equation}\label{eq:det3}
    \prescript{(3)}{}{e}\equiv\det(e^\alpha_i)=\frac{1}{6}\epsilon_{\alpha\beta\gamma}\epsilon^{ijk}e^\alpha_ie^\beta_je^\gamma_k.
\end{equation}

By applying the ADM formalism, the Einstein--Hilbert Lagrangian density in the Palatini formalism can be expressed as follows:
\begin{equation}\label{eq:L_ADM}
    \begin{split}
        \mathcal{L}_\textup{EH}&=\frac{M_\textup{P}^2}{2}ee^\mu_ae^\nu_bR_{\mu\nu}^{ab}=M_\textup{P}^2ee^{[\mu}_ae^{\nu]}_b(\partial_\mu\omega_\nu^{ab}+\omega^a_{c\mu}\omega_\nu^{cb})\\
        &=\bar{\pi}^i_{ab}\dot{\omega}_i^{ab}-N\mathcal{H}_\perp-N^i\mathcal{H}_i-\omega_0^{ab}\mathcal{J}_{ab}-\partial_i(\bar{\pi}^i_{ab}\omega_0^{ab}),
    \end{split}
\end{equation}
with
\begin{subequations}\label{eq:def_Hs}
    \begin{align}
        \bar{\pi}^i_{ab}&\equiv M_\textup{P}^2\prescript{(3)}{}{e}e^{\tilde{i}}_{[a}n_{b]},\\
        \mathcal{H}_\perp&\equiv\frac{M_\textup{P}^2}{2}\prescript{(3)}{}{e}e^{\tilde{i}}_be^{\tilde{j}}_aR^{ab}_{ij}=-\frac{M_\textup{P}^2}{2}\prescript{(3)}{}{e}\prescript{(3)}{}{R},\\
        \mathcal{H}_i&\equiv M_\textup{P}^2\prescript{(3)}{}{e}n_be^{\tilde{j}}_aR^{ab}_{ij}=\bar{\pi}^j_{ab}R^{ab}_{ij},\\
        \mathcal{J}_{ab}&\equiv-\partial_i\bar{\pi}^i_{ab}-2\omega^c_{[a\lvert i\rvert}\bar{\pi}^i_{b]c},
    \end{align}
\end{subequations}
where it was introduced the notation of adding a tilde on every contravariant spatial index $i,j,k$ etc.\ that was raised with the three-dimensional inverse metric $h^{ij}$ rather than the four-dimensional one $g^{\mu\nu}$. So, for example, $e^{\tilde{i}}_a\equiv\eta_{ab}h^{ij}e^b_j$ and $e^i_a=\eta_{ab}g^{i\mu}e^b_\mu$. In particular, using the identities \eqref{eq:e0a-ea0}, one finds that
\begin{equation}\label{eq:tilde_notation}
    e^i_a=e^{\tilde{i}}_a+\frac{N^i}{N}n_a=e^{\tilde{i}}_a-N^ie^0_a.
\end{equation}
The expression \eqref{eq:L_ADM} contains the following constraints: $\mathcal{H}_\perp$ is related to the generator of time evolution; $\mathcal{H}_i$ generates three-dimensional spatial diffeomorphisms; $\mathcal{J}_{ab}$ is the generator of local Lorentz transformations. The form of the Einstein--Hilbert Lagrangian density in the ADM formalism is not unique. The form \eqref{eq:L_ADM} is the same as the one presented in Ref.\ \cite{distefano:canonical}, which differs from the one studied in Ref.\ \cite{castellani:tetrad} because of an integration by parts. Such difference results in a different set of nonzero conjugate momenta. In the present case, the only nonzero conjugate momentum is the one related to $\omega_i^{ab}$. In Ref.\ \cite{castellani:tetrad}, on the other hand, the only nonzero conjugate momentum is the one related to $e^a_i$. As discussed after Eq.\ \eqref{eq:pqs}, in the theory of emergent gravity under examination the only nonzero conjugate momentum after the SSB is effectively the one related to $A_i^{ab}\xrightarrow{SSB}\omega_i^{ab}$. Therefore, the SSB mechanism appears to naturally select one of the two possible scenarios for the Hamiltonian formulation of the gravitational interaction, assigning a more fundamental role to the (spatial) spin connection rather than the (spatial) tetrads. More explicitly, the only nonzero conjugate momentum of the Lagrangian density \eqref{eq:L_ADM} is
\begin{equation}
    \pi^i_{ab}=\frac{\delta\mathcal{L}_\textup{EH}}{\delta\dot{\omega}_i^{ab}}=\bar{\pi}^i_{ab}.
\end{equation}
From this follows that the corresponding Hamiltonian is
\begin{equation}\label{eq:H_ADM}
    \begin{split}
        H_\textup{ADM}&=\int\mathcal{H}_\textup{ADM}\,d^3x\\
        &=\int[N\mathcal{H}_\perp+N^i\mathcal{H}_i+\omega_0^{ab}\mathcal{J}_{ab}+\partial_i(\bar{\pi}^i_{ab}\omega_0^{ab})]\,d^3x.
    \end{split}
\end{equation}
Observe that the Hamiltonian $H_\textup{ADM}$ is not polynomial in all the fields because of the presence of inverse (spatial) tetrads in the quantities \eqref{eq:def_Hs} -- this is also a consequence of the respective Lagrangian density being non-polynomial.

Since the unit normal $n^a$ is orthogonal to all spatial tetrads $e^a_i$ (that is $n_ae^a_i=0$), it can be conveniently expressed as
\begin{equation}\label{eq:n_0}
    n_a\equiv\bar{n}\epsilon_{abcd}\epsilon^{0ijk}e^b_ie^c_je^d_k,\qquad\bar{n}=-\frac{Ne^0_0}{6\prescript{(3)}{}{e}}.
\end{equation}
The normalisation factor $\bar{n}$ can be found, using the identity \eqref{eq:det3}, from the following computation:
\begin{equation*}
    n_0=\bar{n}\epsilon_{0\alpha\beta\gamma}\epsilon^{0ijk}e^\alpha_ie^\beta_je^\gamma_k=6\bar{n}\prescript{(3)}{}{e}=e^\mu_0n_\mu=-Ne^0_0.
\end{equation*}
With this alternative expression for the normal vector to the spatial hypersurfaces as well as the formulae \eqref{eq:e0a-ea0} and \eqref{eq:tilde_notation}, the conjugate momentum $\pi^i_{ab}$ can be recast as
\begin{equation}
    \begin{split}
        \pi^i_{ab}&=\bar{\pi}^i_{ab}=M_\textup{P}^2\prescript{(3)}{}{e}(e^i_{[a}n_{b]}+N^ie^0_{[a}n_{b]})\\
        &=M_\textup{P}^2\prescript{(3)}{}{e}\biggl(\bar{n}e^i_{[a}\epsilon_{b]cde}\epsilon^{0jkl}e^c_je^d_ke^e_l\cancel{-\frac{N^i}{N}n_{[a}n_{b]}}\biggr)\\
        &=-\frac{M_\textup{P}^2Ne_0^0}{6}e^i_{[a}\epsilon_{b]cde}\epsilon^{0jkl}e^c_je^d_ke^e_l.
    \end{split}
\end{equation}
For comparison, the conjugate momentum $\Pi_{ab}^i$ in the spontaneously broken phase, according to the Eqs.\ \eqref{eq:M_P^2} and \eqref{eq:SSB_A}, is
\begin{equation}
    \Pi^i_{ab}=-\frac{M_\textup{P}^2}{4}\epsilon_{abcd}\epsilon^{0ijk}e^c_je^d_k.
\end{equation}
If, again, the tetrads are assumed to be invertible, then the following identity holds true:
\begin{equation}\label{eq:id_momenta}
    -\frac{1}{3}e^i_{[a}\epsilon_{b]cde}\epsilon^{0jkl}e^c_je^d_ke^e_l=\frac{1}{2}\epsilon_{abcd}\epsilon^{0ijk}e^c_je^d_k.
\end{equation}
The result \eqref{eq:id_momenta} can be used to show that the conjugate momentum to $\omega_i^{ab}$ in $\mathcal{H}_\textup{ADM}$ and the conjugate momentum to $A_i^{ab}\xrightarrow{SSB}\omega_i^{ab}$ in $\mathcal{H}_\textup{W}$ after the SSB are equal if and only if $Ne_0^0=-1$, i.e.
\begin{equation}\label{eq:gauge}
    \Pi^i_{ab}=\pi^i_{ab}=\bar{\pi}^i_{ab}\Leftrightarrow e_0^0=-\frac{1}{N}\Leftrightarrow n_0=1,
\end{equation}
with the last implication following from the identity \eqref{eq:e0a-ea0}. Recalling the normalisation condition
\begin{equation*}
    -1=n_\mu n^\mu=n_an^a=-n_0^2+n_1^2+n_2^2+n_3^2,
\end{equation*}
the gauge-fixing condition \eqref{eq:gauge} for the ADM formalism can be seen to coincide with the time gauge of Loop Quantum Gravity in the formulation with Ashtekar variables \cite{thiemann:modern}:
\begin{equation*}
    n_0=1\Rightarrow n_\alpha=0\Rightarrow e^0_\alpha=0,
\end{equation*}
which in turn implies that
\begin{equation}\label{eq:gauge-fixing}
    \begin{split}
        n_a=\delta_a^0,&\qquad n^a=-\delta^a_0,\\
        e^0_a=-\frac{\delta^0_a}{N},\qquad e^0_\mu&=-N\delta^0_\mu,\qquad e^i_0=\frac{N^i}{N}.
    \end{split}
\end{equation}
Therefore, the Hamiltonian of the Einstein--Cartan theory (plus the cosmological constant term) in the ADM formalism is equivalent to that of a theory of emergent gravity {\it à la} Wilczek after the SSB if and only if the gauge freedom in the former is exploited in order to select the time gauge \eqref{eq:gauge-fixing}:
\begin{equation}\label{eq:correspondence_Hs}
    \begin{split}
        \mathcal{H}_\textup{ADM}=\pi^i_{ab}&\dot{\omega}_i^{ab}-(\mathcal{L}_\textup{EH}+\mathcal{L}_\Lambda),\\
        \mathcal{H}_\textup{W}\xrightarrow{SSB}\Pi^i_{ab}&\dot{\omega}_i^{ab}-(\mathcal{L}_\textup{EH}+\mathcal{L}_\Lambda).
    \end{split}
\end{equation}
From a different perspective, the Hamiltonian formulation of the gravitational interaction {\it à la} Wilczek naturally fixes the time gauge of the ADM formalism via the SSB that leads to the emergence of a metric structure in spacetime, without ever requiring a foliation into spatial hypersurfaces. This result is consistent with the fact that that {\it à la} Wilczek can be recast as an extended BF theory with a particular constraint, just like GR \cite{thiemann:modern}. We will expand on this observation in Sec.\ \ref{sec:6}.

The correspondence \eqref{eq:correspondence_Hs} can be analysed more explicitly by studying the Hamiltonian \eqref{eq:H_W(broken)} term by term. The first two terms of \eqref{eq:H_W(broken)} correspond to the third term of \eqref{eq:H_ADM}:
\begin{equation}
    \begin{split}
        \frac{M_\textup{P}^2}{4}&\epsilon_{abcd}\epsilon^{0ijk}[\omega_0^{ab}\partial_i(e^c_je^d_k)+2\omega_{e0}^a\omega_i^{eb}e^c_je^d_k]\\
        &=-\omega_0^{ab}\partial_i\bar{\pi}^i_{ab}-2\omega_0^{ab}\omega^c_{ai}\bar{\pi}^i_{bc}=\omega_0^{ab}\mathcal{J}_{ab}.
    \end{split}
\end{equation}
Given the identity
\begin{equation*}
    \epsilon_{abcd}\epsilon^{0ijk}e_0^ae_i^bR^{cd}_{jk}=2ee^i_be^j_aR^{ab}_{ij},
\end{equation*}
with the use of the formula \eqref{eq:tilde_notation} one can show that in the time gauge \eqref{eq:gauge-fixing} the fourth term of \eqref{eq:H_W(broken)} reproduces the first two terms of \eqref{eq:H_ADM} because
\begin{equation}
    N\mathcal{H}_\perp+N^i\mathcal{H}_i=\frac{M_\textup{P}^2}{2}ee^i_be^j_aR^{ab}_{ij}.
\end{equation}
The fifth term of \eqref{eq:H_W(broken)} is exactly the fourth term of \eqref{eq:H_ADM}:
\begin{equation}
    -\frac{M_\textup{P}^2}{4}\epsilon_{abcd}\epsilon^{0ijk}\partial_i(\omega_0^{ab}e^c_je^d_k)=\partial_i(\bar{\pi}^i_{ab}\omega_0^{ab}).
\end{equation}
Finally, using the definition \eqref{eq:lambda} one finds that the third term of \eqref{eq:H_W(broken)} provides the contribution of the cosmological constant:
\begin{equation}
    \mp M_\textup{P}^2m^2\epsilon_{abcd}\epsilon^{0ijk}e^a_0e^b_ie^c_je^d_k=\pm6M_\textup{P}^2m^2e=M_\textup{P}^2\Lambda e.
\end{equation}
This can be included in the ADM analysis by adding the cosmological constant term $\mathcal{L}_\Lambda$ to $\mathcal{L}_\textup{EH}$, from which follows that
\begin{equation*}
    \mathcal{H}_\Lambda=M_\textup{P}^2\Lambda e.
\end{equation*}

Therefore, the SSB of the pre-geometric theory of gravity under scrutiny can successfully recover all results of the canonical Hamiltonian formulation of GR. As a consequence, its formalism is fully compatible with that of Loop Quantum Gravity as well. For instance, it is intriguing to note that the pre-geometric theory can recover Ashtekar's electric field variables in a natural way, the SSB of the conjugate momenta to the gauge fields $A_i^{A0}$ being given by
\begin{equation}
    \begin{split}
        \Pi^i_{A0}\xrightarrow{SSB}\Pi^i_{a0}&=\Pi^i_{\alpha0}=\frac{M_\textup{P}^2}{4}\epsilon_{0\alpha\beta\gamma}\epsilon^{0ijk}e^\beta_je^\gamma_k\\
        &=\frac{M_\textup{P}^2}{2}\tilde{E}^i_\alpha=\frac{M_\textup{P}^2}{2}\sqrt{h}E^i_\alpha=\frac{M_\textup{P}^2}{2}\prescript{(3)}{}{e}e^{\hat{i}}_\alpha,
    \end{split}
\end{equation}
where $E^i_\alpha\equiv e^{\hat{i}}_\alpha$ are the triads and $\tilde{E}^i_\alpha\equiv\sqrt{h}E^i_\alpha$ the densitised triads or the so-called ``electric field'' \cite{thiemann:modern}.

\section{Dirac's Hamiltonian analysis for the Wilczek theory}\label{sec:4}
The total Lagrangian density \eqref{eq:lagrangian} is degenerate because it is a sum of terms which are linear in the `velocities' of the pre-geometric fields. Therefore, to complete the present Hamiltonian analysis, one must apply Dirac's algorithm for constrained systems or gauge theories \cite{dirac:hamiltonian,golovnev:role}. The expressions \eqref{eq:mom_phi} and \eqref{eq:mom_A} of the conjugate momenta on the constraint surface in phase space can be replaced with the following shorthands:
\begin{align}
    \begin{split}
        \Pi_A&(\phi,A)\equiv2\epsilon_{ABCDE}\epsilon^{0ijk}\nabla_k\phi^D\phi^E[k_\textup{W}F_{ij}^{BC}\\
        &-2\sign(J)k_\textup{SSB}v^{-4}\nabla_i\phi^B\nabla_j\phi^C(\phi^2\mp v^2)^2],
    \end{split}\\
    \Pi^\lambda_{AB}&(\phi,A)\equiv2k_\textup{W}\epsilon_{ABCDE}\epsilon^{0\lambda jk}\nabla_j\phi^C\nabla_k\phi^D\phi^E\nonumber;
\end{align}
in particular,
\begin{subequations}
    \begin{align}
        \Pi^i_{AB}&(\phi,A)=2k_\textup{W}\epsilon_{ABCDE}\epsilon^{0ijk}\nabla_j\phi^C\nabla_k\phi^D\phi^E,\\
        \Pi^0_{AB}&(\phi,A)=0.
    \end{align}
\end{subequations}
The three primary constraints of the theory are then
\begin{subequations}\label{eq:primary-constraints}
    \begin{align}
        Z_A&\equiv\Pi_A-\Pi_A(\phi,A)\approx0,\\
        Z^i_{AB}&\equiv\Pi^i_{AB}-\Pi^i_{AB}(\phi,A)\approx0,\\
        Z^0_{AB}&\equiv\Pi^0_{AB}\approx0,
    \end{align}
\end{subequations}
where the symbol $\approx$ denotes a weak equality on the constraint surface. The total Hamiltonian density is obtained from the canonical one of Eq.\ \eqref{eq:H_W(unbroken)} by adding the primary constraints \eqref{eq:primary-constraints} as Lagrange multipliers. Neglecting the boundary term in what follows, the total Hamiltonian density is thus
\begin{equation*}
    \begin{split}
        \mathcal{H}&=-A_0^{AB}[\partial_i\Pi^i_{AB}(\phi,A)+2\Pi^i_{BC}(\phi,A)A_{Ai}^C\\
        &+\eta_{BC}\Pi_A(\phi,A)\phi^C]+\lambda^AZ_A+\lambda_i^{AB}Z^i_{AB}+\lambda_0^{AB}Z^0_{AB},
    \end{split}
\end{equation*}
where $\lambda^A$, $\lambda_i^{AB}$ and $\lambda_0^{AB}$ are arbitrary coefficients. The conservation in time of the primary constraint $Z^0_{AB}$ yields the following secondary constraint:
\begin{equation}
    \begin{split}
        \dot{Z}^0_{AB}&=\{Z^0_{AB},H\}=\partial_i\Pi^i_{AB}(\phi,A)\\
        &+2\Pi^i_{BC}(\phi,A)A_{Ai}^C+\eta_{BC}\Pi_A(\phi,A)\phi^C\approx0.
    \end{split}
\end{equation}
As a consequence, the total Hamiltonian density becomes
\begin{equation}\label{eq:total-H}
    \begin{split}
        \mathcal{H}&=\!-\!A_0^{AB}\dot{Z}^0_{AB}\!+\!\lambda^AZ_A\!+\!\lambda_i^{AB}Z^i_{AB}\!+\!\lambda_0^{AB}Z^0_{AB}\!+\!\tilde{\lambda}_0^{AB}\dot{Z}^0_{AB}\\
        &\equiv\lambda^AZ_A+\lambda_i^{AB}Z^i_{AB}+\lambda_0^{AB}Z^0_{AB}+\tilde{\lambda}_0^{AB}\dot{Z}^0_{AB},
    \end{split}
\end{equation}
where in the last step the two terms proportional to $\dot{Z}^0_{AB}$ were combined by redefining the Lagrange multiplier $\tilde{\lambda}_0^{AB}$. Note that the field $A_0^{AB}$ drops out of the total Hamiltonian density, hence it is a gauge degree of freedom and the coefficient $\lambda_0^{AB}$ of its corresponding primary constraint $Z_{AB}^0$ remains undetermined. The conservation in time of the primary constraints $Z^i_{AB}$ and $Z_A$ does not yield additional secondary constraints. Rather, it fixes implicitly the form of two Lagrange multipliers, say $\lambda_i^{AB}$ and $\lambda^A$ in terms of $\tilde{\lambda}_0^{AB}$, as follows:
\begin{equation}\label{eq:dot-Z^i_AB}
    \begin{split}
        \dot{Z}^i_{AB}&=\{Z^i_{AB},H\}=\int[\lambda^C\{Z^i_{AB}(x),Z_C(y)\}\\
        &+\tilde{\lambda}_0^{CD}\{Z^i_{AB}(x),\dot{Z}^0_{CD}(y)\}]\,d^3y\approx0,
    \end{split}
\end{equation}
\begin{equation}\label{eq:dot-Z_A}
    \begin{split}
        \dot{Z}_A&=\{Z_A,H\}=\int[\lambda_i^{CD}\{Z_A(x),Z^i_{CD}(y)\}\\
        &+\tilde{\lambda}_0^{CD}\{Z_A(x),\dot{Z}^0_{CD}(y)\}]\,d^3y\approx0.
    \end{split}
\end{equation}
Any linear combination of these last two expressions vanishes weakly on the constraint surface, in particular
\begin{equation}\label{eq:result1}
    \lambda_i^{AB}\{Z^i_{AB},H\}+\lambda^A\{Z_A,H\}\approx0.
\end{equation}
Making use of the Eqs.\ \eqref{eq:dot-Z^i_AB} and \eqref{eq:dot-Z_A}, such linear combination can be recast more explicitly as
\begin{equation}\label{eq:result2}
    \begin{split}
         \lambda_i^{AB}&\{Z^i_{AB},H\}+\lambda^A\{Z_A,H\}\\
         &=\tilde{\lambda}_0^{CD}\int[\lambda_i^{AB}\{Z^i_{AB}(x),\dot{Z}^0_{CD}(y)\}\\
         &+\lambda^A\{Z_A(x),\dot{Z}^0_{CD}(y)\}]\,d^3y.
    \end{split}
\end{equation}
From the Eqs.\ \eqref{eq:result1} and \eqref{eq:result2} follows that
\begin{equation}\label{eq:result3}
    \int[\lambda_i^{AB}\{Z^i_{AB}(x),\dot{Z}^0_{CD}(y)\}+\lambda^A\{Z_A(x),\dot{Z}^0_{CD}(y)\}]\,d^3y\approx0.
\end{equation}
The result \eqref{eq:result3} allows to show that the only secondary constraint of the theory is automatically conserved in time:
\begin{equation}
    \begin{split}
        \ddot{Z}^0_{CD}&=\{\dot{Z}^0_{CD},H\}=-\int[\lambda_i^{AB}\{Z^i_{AB}(y),\dot{Z}^0_{CD}(x)\}\\
        &+\lambda^A\{Z_A(y),\dot{Z}^0_{CD}(x)\}]\,d^3y\approx0.
    \end{split}
\end{equation}
The total Hamiltonian $H$ can thus be expressed, using the form \eqref{eq:total-H} for $\mathcal{H}$, as a linear combination of all the primary and secondary constraints of the theory and it is a first-class constraint as well.

To sum up, we have four constraints: three primary constraints $Z_{AB}^0$, $Z_{AB}^i$ and $Z_A$ and one secondary constraint $\dot{Z}_{AB}^0$. Of these, $Z_{AB}^0$ is first-class while the other three are second-class (with their respective Poisson brackets involving complicated expressions of the pre-geometric fields). Only two second-class constraints are independent, because the total Hamiltonian is a linear combination of all the constraints and a first-class constraint itself. The only Lagrange multipliers that are undetermined in the final form of the total Hamiltonian are $\lambda_0^{AB}$ and $\tilde{\lambda}_0^{AB}$. We can now proceed to compute the number of degrees of freedom of the theory. There are 90 dynamical variables (10 $A_0^{AB}$, 30 $A_i^{AB}$, 5 $\phi^A$, 10 $\Pi_{AB}^0$, 30 $\Pi_{AB}^i$ and 5 $\Pi_A$), 20 gauge choices (-10 $A_0^{AB}$ and -10 $\Pi_{AB}^0$), 10 independent first-class constraints (10 $Z_{AB}^0$), 44 independent second-class constraints (30 $Z_{AB}^i$, 5 $Z_A$, 10 $\dot{Z}_{AB}^0$ and -1 $H$). Therefore, the number of degrees of freedom is given by
\begin{equation*}
    \begin{split}
        2&\#(\textup{degrees of freedom})\\
        &=\#(\text{dynamical variables})-\#(\text{gauge choices})\\
        &-2\#(\text{first-class constraints})-\#(\text{second-class constraints})\\
        &=90-20-20-44=6.
    \end{split}
\end{equation*}
The three degrees of freedom of the theory are compatible with those of a massless graviton and a massive scalar field, which is $\phi^5\equiv\rho$, just like in a scalar-tensor \emph{metric} theory of gravity.

Unlike other stability or Hamiltonian analyses of gravitational theories appearing in the literature (see, for example, Refs.\ \cite{jiménez:minkowski,golovnev:nontrivial,golovnev:hamiltonian,golovnev:adm,golovnev:bimetric}), our Hamiltonian analysis of the Wilczek theory is independent of any spacetime background. This is due to the peculiar feature of the pre-geometric spacetime under exam, which is endowed with topological and differential structures, but \emph{not} also a metric structure like in GR and extensions or modifications thereof. In other words, no background need ever be specified for the spacetime of a pre-geometric theory in a general analysis -- not even, say, a Minkowski metric.

\section{Pre-geometric Wheeler--DeWitt equation}\label{sec:5} 
Having found an expression for the total Hamiltonian of the system, we can formally write down a generalisation of the Wheeler--DeWitt equation \cite{dewitt:canonical}:
\begin{equation}\label{eq:W-DW}
    \hat{\mathcal{H}}|\Psi\rangle=0,
\end{equation}
where $\hat{\mathcal{H}}$ is the Hamiltonian operator obtained from the Eq.\ \eqref{eq:total-H} and $|\Psi\rangle$ is a state of the Hilbert space of all the pre-geometric field configurations. In the case of the Wheeler--DeWitt equation of the ADM formalism, the three-dimensional metric $h_{ij}$ is a function of the three spatial coordinates of the spatial hypersurfaces. In the case of the pre-geometric Wheeler--DeWitt equation \eqref{eq:W-DW}, instead, the pre-geometric fields $A_\mu^{AB}$ and $\phi^A$ are functions of all four spacetime coordinates of the spacetime manifold, given that no foliation of spacetime is employed. In the `pre-metric' representation of the pre-geometric superspace, the state $|\Psi\rangle$ is thus a wave functional $\Psi[A_\lambda^{AB}(x^\mu),\phi^C(x^\nu)]$ (instead of $\Psi[h_{ij}(x^k)]$). In this representation, the operator $\hat{\mathcal{H}}$ is obtained by promoting the pre-geometric fields to operators as follows:\footnote{The correspondence rule gives rise to operator ordering ambiguities in the definition of the Hamiltonian operator $\hat{\mathcal{H}}$.}
\begin{align}
    \hat{A}_\mu^{AB}(x)&\rightarrow A_\mu^{AB}(x),\qquad\Pi^\mu_{AB}(x)\rightarrow-i\frac{\delta}{\delta A_\mu^{AB}(x)},\\
    \hat{\phi}^A(x)&\rightarrow\phi^A(x),\qquad\Pi_A(x)\rightarrow-i\frac{\delta}{\delta\phi^A(x)}.
\end{align}

The fact that the Hamiltonian \eqref{eq:total-H} is polynomial in the fields and their conjugate momenta, as noticed previously, is a particularly promising feature for the quantisation scheme, unlike what happens for the canonical quantisation of gravity in the ADM formalism. Moreover, in the pre-geometric theory the problem of time assumes a different connotation: time derivatives are still absent from the Hamiltonian constraint (as they should be due to diffeomorphism invariance), but the pre-geometric fields are functions of time as well, other than position. Therefore, the time coordinate \emph{does} play a role in the pre-geometric Wheeler--DeWitt equation. The correspondence \eqref{eq:correspondence_Hs}, according to which the time gauge yields the result $\mathcal{H}_\textup{W}\xrightarrow{SSB}\mathcal{H}_\textup{ADM}$, implies that the pre-geometric Wheeler--DeWitt equation reduces to the metric Wheeler--DeWitt equation of the ADM formalism after the SSB. This means that the SSB must be responsible also for reducing the pre-geometric superspace to the metric superspace of the ADM formalism. It then remains puzzling to observe that, once the phase transition for the fundamental gauge symmetry of spacetime happens, time ceases to be relevant in the Hamiltonian analysis of the theory, according to the metric Wheeler--DeWitt equation. This could be an indication of the fact that the pre-geometric framework is necessary for studying the canonical quantisation of classical gravity, as the problem of time seemingly becomes `less problematic' therein. 

\section{Formulation of the pre-geomtric theory in the extended BF formalism}\label{sec:6}
It was shown in previous studies that is possible to recast GR within the language of extended BF theories \cite{celada:bf,baez:introduction,baez:topological,witten:topological,atiyah:topological,ashtekar:review,freidel:topological}. We here applied the same strategy to the (dynamical) pre-geometric theory hitherto studied, recovering it as `emergent' from a topological BF theory. The details of the emergence of the pre-geometric theory will be presented in a forthcoming paper. Here we rather provide a qualitative picture of the relation between the Higgs-like scalar multiplet and the thermal history of the universe, underlining a possible action of the Higgs-like field as a `pre-geometric clock'.  Indeed, the finite temperature dynamics of the Higgs-like field being inextricably intertwined to the realisation (in the sense of emergence) of the theory of gravity, is embedded {\it ab initio} in the pre-geometric theory proposed by Wilczek. In Ref.\ \cite{alexander:higgs} the very same idea was implemented through the Higgs field of the Standard Model, while in our work we will focus on the Higgs-like field of the pre-geometric formulation.\footnote{For further studies connecting the SSB of the Higgs field of the Standard Model to GR, see e.g.\ the Refs.\ \cite{percacci:higgs,westman:gauge,westman:cartan,westman:introduction,dehnen:higgs,alexander:gravitational,hehl:metric,adler:einstein,bekenstein:gravitation,krasnov:spontaneous,chamseddine:unification}.}

We start by briefly reviewing the BF framework and its extension. Given a four-dimensional Lorentzian manifold equipped with a principal $G$-bundle, the action for a BF theory (without cosmological constant term) is
\begin{equation}\label{eq:BF}
    S_\textup{BF}=\int\trace(B\wedge F),
\end{equation}
where the trace is computed over the internal indices of the Lie algebra of $G$. The dynamical fields of the theory are the connection form $A$ for $G$, whose curvature is $F=dA+A\wedge A$, and a two-form $B$ taking values in the adjoint representation of $G$. The equations of motion are the Gauss constraint, as the generator of the gauge symmetries of $G$, and the curvature constraint, which imposes the flatness of the topological configurations. These are respectively
\begin{equation*}
    \mathcal{D}B^{AB}=0,\qquad F^{AB}=0,
\end{equation*}
where $\mathcal{D}$ denotes the covariant derivative with respect to the connection $A$. The action \eqref{eq:BF} is invariant not only under diffeomorphisms and gauge transformations, but also under shifts of the $B$ field of the form
\begin{equation*}
    B^{AB}\rightarrow B^{AB}+\mathcal{D}V^{AB}
\end{equation*}
for any one-form $V$. This symmetry results in the absence of local degrees of freedom and, thus, in the topological invariance of the theory.

GR can also be reformulated in the language of a topological BF theory by adopting the Palatini formalism. In this case, the gauge group of the theory is the Lorentz group, i.e.\ $G=SO(1,3)$. The Einstein--Hilbert action, in fact, can be expressed as
\begin{equation}
    S_\textup{EH}=\frac{M_\textup{P}^2}{4}\int\epsilon_{abcd}e^c\wedge e^d\wedge F^{ab}.
\end{equation}
This action is a BF theory like that of Eq.\ \eqref{eq:BF}, but with the requirement that there exist tetrad fields $e^a$ such that the $B$ field assumes the specific form
\begin{equation}\label{eq:simplicity-constraint}
    B_{ab}^{(\textup{EH})}=\frac{M_\textup{P}^2}{4}\epsilon_{abcd}e^c\wedge e^d=\frac{M_\textup{P}^2}{2}{*}(e\wedge e)_{ab},
\end{equation}
where $*$ denotes the Hodge dual with respect to the internal space Minkowski metric $\eta_{ab}$. The condition \eqref{eq:simplicity-constraint} is called ``simplicity constraint'' and is responsible for breaking the topological invariance of the BF theory. It allows to recover GR, which is indeed a theory with two local degrees of freedom, rather than a topological theory. The simplicity constraint can be imposed as a constraint in the action via the introduction of a Lagrange multiplier endowed with appropriate symmetries \cite{thiemann:modern}. This leads to the so-called Plebanski action \cite{plebanski:separation,de-pietri:plebanski},
\begin{equation}\label{eq:plebanski}
    S_\textup{P}=\int(B_{ab}\wedge F^{ab}+\Phi_{abcd}B^{ab}\wedge B^{cd})\equiv S_\textup{BF}+S_c^{(\textup{EH})},
\end{equation}
which is an extended BF theory $S_\textup{BF}$ provided with a constraint $S_c^{(\textup{EH})}$ containing the dimensionless Lagrange multiplier $\Phi_{abcd}=-\Phi_{bacd}=-\Phi_{abdc}=\Phi_{cdab}$. The introduction of the constraint $S_c^{(\textup{EH})}$ is a way to recover GR from the BF formalism via the action principle: extremisation of $S_\textup{P}$ with respect to $\Phi_{abcd}$, in fact, imposes the condition
\begin{equation*}
    B^{ab}\wedge B^{cd}=\frac{1}{4!}\epsilon^{abcd}\epsilon_{efgh}B^{ef}\wedge B^{gh}.
\end{equation*}
The four different solutions of this equation realise the implementation of the simplicity constraint \eqref{eq:simplicity-constraint}, which  reduces\footnote{For the Hamiltonian analysis of BF theories and the Plebanski theory, see the Refs.\ \cite{escalante:dirac} and \cite{buffenoir:hamiltonian,durka:hamiltonian} respectively.} $S_\textup{BF}$ to $S_\textup{EH}$ \cite{thiemann:modern,freidel:topological}.

As noted in Sec.\ \ref{sec:3}, the Wilczek action can be recast in the formalism of a BF theory too, with a condition on the $B$ field that is the pre-geometric generalisation of GR's simplicity constraint, i.e.
\begin{equation}\label{eq:pre-geometric-simplicity-constraint}
    B_{AB}^{(\textup{W})}=-2k_\textup{W}\epsilon_{ABCDE}\nabla\phi^C\wedge\nabla\phi^D\phi^E.
\end{equation}
In doing so, in fact, the recovery of the correct metric theory of gravity after the SSB is ensured by the result
\begin{equation}\label{eq:SSB_constraints}
    B_{AB}^{(\textup{W})}\xrightarrow{SSB}B_{ab}^{(\textup{W})}=B_{ab}^{(\textup{EH})}.
\end{equation}

From the perspective of the BF formalism, it is thus the pre-geometric simplicity constraint \eqref{eq:pre-geometric-simplicity-constraint} that `unfreezes' the three dynamical degrees of freedom of the Wilczek action and sets it apart from topological BF theories. \\

The mechanism through which the pre-geometric theory recovers GR in the IR limit is well understood. In particular, after the SSB this provides a new degree of freedom other than those of the massless graviton, i.e.\ that of the scalar field $\phi^5\equiv\rho$. We may then wonder whether such additional degree of freedom can be relevant for obtaining a UV completion of the theory, given that the mass of its excitations is naturally expected to be near the Planck scale \cite{addazi:pre-geometry,meluccio:field}. Indeed, under the assumption that the pre-geometric fields are coupled to the thermal bath of the early universe, the dynamical mechanism enabling this novel regime of the gravitational interaction can be simply the evolution of the Higgs-like field determined by a temperature-dependent effective potential $V_\textup{eff}(T)$. Provided that a suitable pre-geometric kinetic term for the field $\phi^A$ is supplemented as described in Ref.\ \cite{addazi:pre-geometry}, a common way \cite{mukhanov:cosmology} to model the effective potential in the pre-geometric formalism is
\begin{equation}
    \begin{split}
        V_\textup{eff}(T)&=-\mathcal{L}_\textup{SSB}+\lvert J\rvert(V_2T^2\phi^2+V_3T\phi^3)\\
        &=\lvert J\rvert[V_2(T^2-T_c^2)\phi^2+V_3T\phi^3+V_4\phi^4],
    \end{split}
\end{equation}
where $\mathcal{L}_\textup{SSB}$ and $\phi$ are defined in Eq.\ \eqref{eq:potential}, $V_2$, $V_3$ and $V_4$ are constants and $T_c$ is a critical temperature. The values of these constants define the precise shape of the potential $V_\textup{eff}(T)$, which in turn determines the type of phase transition happening at the critical temperature $T_c$: either a discontinuous, first-order phase transition or a smooth, second-order phase transition. In what follows, we do not need to know the values of these constants, but rather seek only a proof of concept. We hence develop preliminary qualitative arguments, which yet shed a clear physical perspective on how the pre-geometric Wilczek theory may emerge from a topological BF one.\\

The total pre-geometric Plebanski (PGP) action $S_\textup{PGP}$ of the theory is nothing but a rewriting of $S_\textup{P}$ with two differences: $SO(1,4)$ or $SO(3,2)$ as the gauge group rather than $SO(1,3)$, and a slightly modified expression for the constraint. Therefore, we write
\begin{equation}\label{eq:BF_gravitational}
    \begin{split}
        S_\textup{PGP}&=\int(B_{AB}\wedge F^{AB}+k_c\epsilon_{ABCDE}B^{AB}\wedge B^{CD}\phi^E )\\
        &\equiv S_\textup{BF}+S_c^{(\textup{W})},
    \end{split}
\end{equation}
where $k_c$ is a constant. The extremisation of $S_\textup{PGP}$ with respect to $\phi^{E}$ imposes the condition
\begin{equation*}
    B^{AB}\wedge B^{CD}=\frac{1}{4!}\epsilon^{ABCDE}\epsilon_{EFGHI}B^{FG}\wedge B^{HI}.
\end{equation*}
The four different solutions of such equation include the pre-geometric simplicity constraint \eqref{eq:pre-geometric-simplicity-constraint}, which  reduces $S_\textup{BF}$ to $S_\textup{W}$, with $S_\textup{W}\xrightarrow{SSB}S_\textup{EH}+S_\Lambda$.\\

Summarising, the simplicity constraint \eqref{eq:simplicity-constraint} or its pre-geometric generalisation \eqref{eq:pre-geometric-simplicity-constraint} reduce a topological BF theory to a gravitational or to a pre-geometric theory, respectively characterised by two or three local degrees of freedom. So far, these two scenarios have been discussed as two distinct and independent possibilities, given that they are implemented via different actions: the simplicity constraint \eqref{eq:simplicity-constraint} is obtained from the Plebanski action \eqref{eq:plebanski}, while the pre-geometric one \eqref{eq:pre-geometric-simplicity-constraint} is obtained from the PGP action \eqref{eq:BF_gravitational}. The key observation is then that only one action is actually needed for obtaining both the pre-geometric theory $S_\textup{W}$ and the metric one $S_\textup{EH}+S_\Lambda$, and that is the PGP action \eqref{eq:BF_gravitational}. In fact, variation with respect to the multiplet of fields $\phi^{E}$ yields the pre-geometric theory $S_\textup{W}$ on-shell, which in turn reduces to the metric theory $S_\textup{EH}+S_\Lambda$ as a result of the dynamical mechanism of SSB. With respect to the work in Ref.~\cite{alexander:higgs}, the advantage of our model is that we avoid the possible instability due to the cosmological constant, or the fine-tuning thereof. This is because the cosmological constant is emergent in the pre-geometric picture and need not be introduced, for example, as a dynamical field related to the Higgs field.\\

The UV completion of the gravitational interaction within this context requires to understand how the scalar multiplet may actually acquire a dynamics, instead of remaining the Lagrangian multiplier of the PGP action. We will present in detail a possible mechanism that enables to unfreeze the additional scalar degree of freedom in a forthcoming study. Here, for this qualitative discussion, we comment on the dynamics of the Higgs-like field when the topological symmetry is already lost, i.e.\ under the assumption that the phase transition from the topological BF phase to the non-topological geometric phase has already occurred. We therefore consider a qualitative temperature-dependent evolution of the Higgs-like field as follows. For temperatures above $T_c$, the effective potential $V_\textup{eff}(T)$ has only one minimum at $\langle\phi^A\rangle=0$. In this qualitative picture, the constraint $S_c^{(\textup{W})}$ would vanish and $S_\textup{PGP}$ would reduce to $S_\textup{BF}$. The critical temperature at which the SSB happens \cite{addazi:pre-geometry} is expected to be close to the Planck scale, i.e.\ $T_c\lesssim M_\textup{P}$. For temperatures below $T_c$, in this exemplified picture, the Higgs-like field would acquire a nonzero v.e.v.\ $\langle\phi^A\rangle$, which would effectively give rise to the Wilczek theory $S_\textup{W}$ as a constrained BF theory, thanks to the pre-geometric simplicity constraint \eqref{eq:pre-geometric-simplicity-constraint}. This simple toy-model discussion clarifies how the relevance of the pre-geometric phase is actually dependent on the shape of the effective potential, and thus on the order of the phase transition that happens at $T_c$. In the case of a second-order phase transition, the Higgs-like field evolves smoothly to reach the v.e.v.\ $\langle\phi^A\rangle=v\delta^A_5$, thus reducing $S_\textup{BF}$ directly to $S_\textup{EH}+S_\Lambda$ via the relation \eqref{eq:SSB_constraints}. In the case of a first-order phase transition, instead, the Higgs-like field can initially remain trapped in a false vacuum state with $\langle\phi^A\rangle\ne0$, which enacts $S_\textup{W}$, until it reaches the true vacuum state with $\langle\phi^A\rangle=v\delta^A_5$, which then yields $S_\textup{EH}+S_\Lambda$. 

\section{Conclusions}\label{sec:7}
We performed a background-independent Hamiltonian analysis of pre-geometric theories of gravity. Our focus has been preferably on the Wilczek theory, rather than the MacDowell--Mansouri theory. In fact the latter cannot be recast as a constrained BF theory with a generalisation of GR's simplicity constraint -- further reasons were discussed in Ref.\ \cite{addazi:pre-geometry}. We showed that the canonical formulation of GR is correctly recovered in the spontaneously broken phase of a pre-geometric gauge theory, and that the respective Hamiltonian analyses agree after the SSB if and only if the time gauge is selected in the ADM formalism. Furthermore, Dirac's algorithm revealed the constraint algebra of the pre-geometric theories and allowed to count the number of their dynamical degrees of freedom.

The result of three degrees of freedom for both pre-geometric theories is consistent with the fact that, after the SSB, they reproduce classical gravity, in the form of the massless spin-2 graviton, coupled to an additional scalar field which is the excitation of the pre-geometric Higgs-like field with a nonzero v.e.v. In other words, these pre-geometric theories of gravity reduce to scalar-tensor \emph{metric} theories of gravity after the SSB. The result about the degrees of freedom also shows that the $\sigma$ fields of the previous study \cite{addazi:pre-geometry} must not be understood as independent fields on-shell, rather as possible auxiliary fields with no bearing on dynamics.\footnote{This situation is, in a broader sense, reminiscent of the auxiliary fields introduced in SUSY, which would suggest that a kinetic term for the $\sigma$ fields is irrelevant.}  
In the Higgs mechanism of the pre-geometric theories, the degrees of freedom associated to a conceivable mass of the tetrad fields are suppressed by the additional `gravitational' constraints of the theory, like the metric-compatibility condition. Therefore, this result eliminates the possible problem of the tachyonic instabilities highlighted in the previous analysis \cite{addazi:pre-geometry}. Not only, it also suggests that negative-norm states are unphysical in these theories, which is surprising given the non-compactness of their gauge groups. As a consequence, one would expect to be able to find suitable conditions in the quantisation scheme in order to remove negative-norm states from the Fock space, hence guaranteeing the preservation of unitarity.

The knowledge of the Hamiltonian for a pre-geometric formulation of the gravitational interaction allows to write down a generalisation of the Wheeler--DeWitt equation. This insight deserves further investigation, especially for understanding issues like the nature of the pre-geometric superspace, operator ordering ambiguities in the quantisation scheme, the problem of time, etc. In general, the Hamiltonian analysis is a fundamental tool to investigate the degrees of freedom of any theory of gravity -- see e.g.\ Refs.\ \cite{capozziello:hamiltonian,bajardi:hamilton,bajardi:primary}.

Finally, we drew a relation between the pre-geometric theory we have investigated and BF formulations of gravity and pre-geometric theories. In particular, we suggested that the evolution of the Higgs-like field in the early Universe could act as a `pre-geometric thermal time' in determining the specific theory of gravity that is realised in any given energy regime, providing an effective pre-geometric time deparametrisation. Below a certain critical energy, which would be expected to be just below that of the Planck scale, the pre-geometric theory reduces to the metric theory of classical gravity via the mechanism of SSB. In the ultra-high-energy regime, instead, the pre-geometric theory becomes topological, because of the vanishing of the constraint in the BF formulation, thus rendering gravity, in a sense, trivial. This symmetry restoration has the dramatic consequence of depriving the gravitational interaction of all local degrees of freedom, as well as invalidating its geometric meaning. Therefore, this scenario could represent the analogue of a \emph{trivial} UV fixed point for asymptotic safety, but in a very different framework which is that of topological BF theories. The main interest in these formulations of gravity is that a great deal is known about the quantisation of topological BF theories \cite{thiemann:modern,baez:introduction,baez:topological}. In particular, they can provide examples of background independence at the quantum level in the form of topological quantum field theories \cite{witten:topological,atiyah:topological,barrett:lorentzian}. For this reason, we argue that the dynamical mechanism realised by the Higgs-like field may provide a UV completion of gravity at all energy scales. In a forthcoming work, we will delve into this subject through a more detailed analysis.

\appendix*
\section{Dirac's Hamiltonian analysis for the MacDowell--Mansouri theory}\label{sec:appendix}
\setcounter{equation}{0}
\renewcommand{\theequation}{A\arabic{equation}}
The Hamiltonian analysis for the MacDowell--Mansouri theory follows the same steps of that for the Wilczek theory and reaches the same qualitative conclusions.

The Lagrangian density considered in this appendix is the sum of the potential term \eqref{eq:potential} for the realisation of the SSB and the MacDowell--Mansouri Lagrangian density \cite{macdowell:unified},
\begin{equation}
    \mathcal{L}_\textup{MM}=k_\textup{MM}\epsilon_{ABCDE}\epsilon^{\mu\nu\rho\sigma}F_{\mu\nu}^{AB}F_{\rho\sigma}^{CD}\phi^E,
\end{equation}
where $k_\textup{MM}$ is a nonzero constant. In this case, the conjugate momenta to the pre-geometric fields $\phi^A$ and $A_\mu^{AB}$ are respectively
\begin{align}\label{eq:mom_phi_2}
    \begin{split}
        \Pi_E=\frac{\delta(\mathcal{L}_\textup{MM}+\mathcal{L}_\textup{SSB})}{\delta\dot{\phi}^E}=&-4\sign(J)k_\textup{SSB}v^{-4}\epsilon_{ABCDE}\epsilon^{0ijk}\\
        &\cdot\nabla_i\phi^A\nabla_j\phi^B\nabla_k\phi^C\phi^D(\phi^2\mp v^2)^2,
    \end{split}\\
    \Pi^\lambda_{DE}=\frac{\delta(\mathcal{L}_\textup{MM}+\mathcal{L}_\textup{SSB})}{\delta\dot{A}_\lambda^{DE}}=&4k_\textup{MM}\epsilon_{ABCDE}\epsilon^{0\lambda jk}F_{jk}^{AB}\phi^C\nonumber;
\end{align}
in particular,
\begin{subequations}\label{eq:mom_A_2}
    \begin{align}
        \Pi^i_{DE}&=4k_\textup{MM}\epsilon_{ABCDE}\epsilon^{0ijk}F_{jk}^{AB}\phi^C,\\
        \Pi^0_{DE}&=0.
    \end{align}
\end{subequations}
Observe that, unlike what happens for the theory $\mathcal{L}_\textup{W}$, in this case $\Pi_A=0$ if $\mathcal{L}_\textup{SSB}$ is not included in the Hamiltonian analysis. In terms of the conjugate momenta, the total Lagrangian density can be recast as
\begin{equation}
    \mathcal{L}_\textup{MM}+\mathcal{L}_\textup{SSB}=\Pi^i_{AB}F_{0i}^{AB}+\Pi_A\nabla_0\phi^A,
\end{equation}
just like Eq.\ \eqref{eq:lagrangian}. The corresponding Hamiltonian density $\mathcal{H}_\textup{MM}$ and Hamiltonian $H_\textup{MM}$ have the same (implicit) expressions as $\mathcal{H}_\textup{W}$ in Eq.\ \eqref{eq:H-density_W} and $H_\textup{W}$ in Eq.\ \eqref{eq:H_W(unbroken)} respectively, only the explicit expressions for the conjugate momenta being different.

In this case, the SSB of the vacuum state of the theory yields
\begin{equation}
    \begin{split}
        \mathcal{L}_\textup{MM}\xrightarrow{SSB}&\mp4k_\textup{MM}vm^2\epsilon_{abcd}\epsilon^{\mu\nu\rho\sigma}e_\mu^ae_\nu^bR_{\rho\sigma}^{cd}\\
        &-96k_\textup{MM}vm^4e-4k_\textup{MM}ve\mathcal{G},
    \end{split}
\end{equation}
where the Euler characteristic of the spacetime manifold is given by
\begin{equation*}
    \mathcal{G}\equiv R^2-4R_{\mu\nu}R^{\mu\nu}+R_{\mu\nu\rho\sigma}R^{\mu\nu\rho\sigma}.
\end{equation*}
By identifying the reduced Planck mass and the cosmological constant this time respectively as
\begin{subequations}
    \begin{align}
        M_\textup{P}^2&\equiv\pm32k_\textup{MM}vm^2,\label{eq:M_P^2_2}\\
        \Lambda&\equiv\pm3m^2=\frac{3M_\textup{P}^2}{32k_\textup{MM}v},
    \end{align}
\end{subequations}
the SSB of the MacDowell--Mansouri Lagrangian density can be seen to reproduce the gravitational Lagrangian density in the form
\begin{equation*}
    \mathcal{L}_\textup{MM}\xrightarrow{SSB}\mathcal{L}_\textup{EH}+\mathcal{L}_\Lambda+\mathcal{L}_\textup{GB},
\end{equation*}
with the last term identifiable as the Guass--Bonnet term:
\begin{equation}\label{eq:GB}
    \mathcal{L}_\textup{GB}\equiv\lambda e\mathcal{G},\qquad\lambda\equiv-4k_\textup{MM}v.
\end{equation}
After the SSB, the conjugate momenta \eqref{eq:mom_phi_2} and \eqref{eq:mom_A_2} become respectively
\begin{equation}
    \Pi_A\xrightarrow{SSB}\Pi_a=0
\end{equation}
and
\begin{subequations}
    \begin{align}
        \Pi^i_{AB}\xrightarrow{SSB}\Pi^i_{ab}&=4k_\textup{MM}v\epsilon_{abcd}\epsilon^{0ijk}F_{jk}^{cd},\label{eq:SSB_A_2}\\
        \Pi^0_{AB}\xrightarrow{SSB}\Pi^0_{ab}&=0.
    \end{align}
\end{subequations}
Note that Eq.\ \eqref{eq:pqs} holds true also in this case, hence the (spatial) spin connection plays a privileged role in the Hamiltonian formulation of this theory as well.

To correctly carry out the comparison between GR in the ADM formalism and the SSB of the MacDowell--Mansouri theory, the Einstein--Hilbert action must be supplemented with the Gauss--Bonnet term. Knowing that the latter term can also be expressed \cite{addazi:pre-geometry} as
\begin{equation*}
    \mathcal{L}_\textup{GB}=-\frac{\lambda}{4}\epsilon_{abcd}\epsilon^{\mu\nu\rho\sigma}R_{\mu\nu}^{ab}R_{\rho\sigma}^{cd}=-\lambda\epsilon_{abcd}\epsilon^{0ijk}R_{0i}^{ab}R_{jk}^{cd},
\end{equation*}
in this case the only nonzero conjugate momentum of the Lagrangian density in the ADM formalism is found to be
\begin{equation}
    \pi^i_{ab}=\frac{\delta(\mathcal{L}_\textup{EH}+\mathcal{L}_\textup{GB})}{\delta\dot{\omega}_i^{ab}}=\bar{\pi}^i_{ab}-\lambda\epsilon_{abcd}\epsilon^{0ijk}R_{jk}^{cd}.
\end{equation}
For comparison, the conjugate momentum $\Pi_{ab}^i$ in the spontaneously broken phase, according to the Eqs.\ \eqref{eq:M_P^2_2}, \eqref{eq:GB} and \eqref{eq:SSB_A_2}, is
\begin{equation}
    \begin{split}
        \Pi^i_{ab}&=-\lambda\epsilon_{abcd}\epsilon^{0ijk}(R_{jk}^{cd}\mp2m^2e_j^ce_k^d)\\
        &=-\lambda\epsilon_{abcd}\epsilon^{0ijk}R_{jk}^{cd}-\frac{M_\textup{P}^2}{4}\epsilon_{abcd}\epsilon^{0ijk}e_j^ce_k^d.
    \end{split}
\end{equation}
The first of the two terms in this expression of $\Pi^i_{ab}$ is the same as the contribution to $\pi^i_{ab}$ coming from the Gauss--Bonnet term. Just like in the case of the Wilczek theory, the second term is equal to $\bar{\pi}^i_{ab}$ if and only if the time gauge is selected in the ADM formalism, i.e.\ the result \eqref{eq:gauge} holds true in the Hamiltonian analysis of the MacDowell--Mansouri theory too.

We can therefore repeat the same comment made for the result \eqref{eq:correspondence_Hs} of the Wilczek theory. The Hamiltonian of the Einstein--Cartan theory (plus the cosmological constant and the Gauss--Bonnet terms) in the ADM formalism is equivalent to that of a theory of emergent gravity {\it à la} MacDowell--Mansouri after the SSB if and only if the gauge freedom in the former is exploited in order to select the time gauge \eqref{eq:gauge-fixing}:
\begin{equation}
    \begin{split}
        \mathcal{H}_\textup{ADM}=\pi^i_{ab}&\dot{\omega}_i^{ab}-(\mathcal{L}_\textup{EH}+\mathcal{L}_\Lambda+\mathcal{L}_\textup{GB}),\\
        \mathcal{H}_\textup{MM}\xrightarrow{SSB}\Pi^i_{ab}&\dot{\omega}_i^{ab}-(\mathcal{L}_\textup{EH}+\mathcal{L}_\Lambda+\mathcal{L}_\textup{GB}).
    \end{split}
\end{equation}

Dirac’s algorithm for the MacDowell--Mansouri theory yields the same qualitative results obtained for the Wilczek theory, the only quantitative difference being the explicit expressions of the conjugate momenta in the constraints. In particular, the counting of the degrees of freedom is the same as that for the Wilczek theory, thus leading to the same result of three degrees of freedom.
\\

{\bf Acknowledgements}.
AA acknowledges the support of the National Science Foundation of China (NSFC) through the grant No.\ 12350410358; the Talent Scientific Research Program of College of Physics, Sichuan University, Grant No.\ 1082204112427; the Fostering Program in Disciplines Possessing Novel Features for Natural Science of Sichuan University, Grant No.\ 2020SCUNL209 and 1000 Talent program of Sichuan province 2021. SC and GM acknowledge the support of Istituto Nazionale di Fisica Nucleare, Sez.\ di Napoli, Iniziative Specifiche QGSKY and MoonLight-2. SC thanks the Gruppo Nazionale di Fisica Matematica (GNFM) of Istituto Nazionale di Alta Matematica (INDAM) for the support. AM acknowledges the support by the NSFC, through the grant No.\ 11875113, the Shanghai Municipality, through the grant No.\ KBH1512299, and by Fudan University, through the grant No.\ JJH1512105. This paper is based upon work from COST Action CA21136 -- Addressing observational tensions in cosmology with systematics and fundamental physics (CosmoVerse), supported by COST (European Cooperation in Science and Technology).

\end{document}